# Distributed Robust Bilinear State Estimation for Power Systems with Nonlinear Measurements

Weiye Zheng, Wenchuan Wu, *Senior Member, IEEE,* Antonio Gomez-Exposito, *Fellow, IEEE*, Boming Zhang, *Fellow, IEEE*

*Abstract*—This paper proposes a fully distributed robust state-estimation (D-RBSE) method that is applicable to multi-area power systems with nonlinear measurements. We extend the recently introduced bilinear formulation of state estimation problems to a robust model. A distributed bilinear state-estimation procedure is developed. In both linear stages, the state estimation problem in each area is solved locally, with minimal data exchange with its neighbors. The intermediate nonlinear transformation can be performed by all areas in parallel without any need of inter-regional communication. This algorithm does not require a central coordinator and can compress bad measurements by introducing a robust state estimation model. Numerical tests on IEEE 14-bus and 118-bus benchmark systems demonstrate the validity of the method.

## I. Introduction

STATE estimation (SE) is conventionally performed at individual regional control center with very limited interaction between control centers. With the deregulation of electricity markets, growing quantities of power are transferred over tie-lines [1]. Meanwhile, to reduce unnecessarily excessive operation cost [2] and hedge the uncertainty of renewable energy generation [3], regional independent system operators (ISO) should coordinate with other ISOs though the interconnected networks. Recently, many regional markets, such as NYISO [4], ISO-NE[5], PJM [6], and MISO [7] are actively developing coordination schemes and procedures. This type of coordination should base on compatible network models for each area, so a fully distributed multi-area state estimation is needed. Since no coordinators exist above the ISOs, this computation framework is intended to solve the compatible real-time states while preserving information privacy of subsystems.

Distributed SE has been extensively studied under decomposition-coordination framework [8]-[10]. Recently, fully distributed SE methods were proposed, which do not need any central coordination. Alternating direction method of multipliers (ADMM) has been used in [11] to formulate a distributed SE for linear system. Although linear measurements can be incorporated using synchronized phasor measurement units (PMUs), their deployment is currently limited and SE still relies significantly on nonlinear measurements. Therefore, distributed SE that can handle nonlinear measurements is of greater value for practical application. The Auxiliary Problem Principle (APP) has been utilized in[12], whereby each agent solves its own sub-problem and communicates only with its neighboring units. An approximate algorithm based on the optimality condition decomposition has been proposed in[13], however, methods of this kind assume local observability and their convergence is not always guaranteed. Note that these early distributed algorithms [11]-[13] do not deal with the non-convexity issue of nonlinear SE. Since convexity is a prerequisite for guaranteed convergence of most distributed algorithms, semidefinite relaxation (SDR) and ADMM are combined in [14] to provide a distributed algorithm for tree-connected control areas with guaranteed convergence.

Bilinear state estimation (BSE) has been proposed in [15]-[18] as an alternative to the conventional SE based on Gauss-Newton method. The burden of traditional iterative linearization process has been significantly relieved by the non-iterative BSE scheme, which decomposes the original nonlinear SE model into two linear stages accompanying a nonlinear transformation with the help of intermediate variables.

In this paper, the BSE proposed in [16][18] is extended to handle multi-area power systems in presence of bad data. Each of the three steps is further decoupled over different areas, yielding a fully distributed robust bilinear state estimation (D-RBSE) with guaranteed convergence thanks to ADMM. For the two linear stages, each area solves its regional SE subproblem, sends the latest boundary states to its neighboring areas, and iterates in this way until convergence; the interleaved nonlinear transformation can be processed within each area in parallel without any need of inter-regional communication. This D-RBSE is applicable for power systems with arbitrary network configuration and nonlinear measurements. It also has higher efficiency and guaranteed convergence compared with existing methods.

The remainder of the paper is organized as follows. In Section II, power system state estimation model is briefly reviewed. Section III describes the robust bilinear state estimation (RBSE). A fully distributed algorithm to solve multi-area RBSE is described in Section IV. Section V details the results of several numerical tests to investigate the performance of D-RBSE. Section VI concludes the paper.

## II. Power System State Estimation

The measurement model for power systems is[19]:
$$z = h(x) + e \qquad (1)$$

Manuscript received XX, 2015. This work was supported in part by the National Key Basic Research Program of China (2013CB228205), in part by the National Science Foundation of China (51477083).

W. Zheng, W. Wu, B. Zhang are with the State Key Laboratory of Power Systems, Department of Electrical Engineering, Tsinghua University, Beijing 100084, China (e-mail: wuwench@tsinghua.edu.cn). A. Gomez-Exposito is with the Department of Electrical Engineering, University of Seville, Seville, Spain (e-mail: age@us.es).

where $h(x)$ denotes a vector of functions describing error-free measurements of state variables $x$, and $e$ denotes the vector of measurement errors, which is generally assumed to be $N(0,\sigma)$ and uncorrelated.

Exactly-known magnitudes (such as zero injection constraints) should be satisfied by the estimators. Since considering these as very accurate measurements with very large weightings will bring about numerical problems [19], such constraints are added explicitly to the estimation model as follows:

$$h_e(x) = z_e \quad (2)$$

where $z_e = 0$ for zero injected power.

The least square SE model can be formulated as

$$\min \quad J(x) = \frac{1}{2}[z-h(x)]^T[z-h(x)]$$
$$\text{s.t.} \quad h_e(x) = z_e \quad (3)$$

In practice, real-time measurements may be corrupted by data contamination, instrument failure and asynchronous meter measurements [19]. In the context of the cyber-physical smart grid, bad data may result not only from unintentional metering faults, but also malicious cyber-attack [20]. In the presence of bad data, a more detailed measurement model is given by[20][21]

$$z = h(x) + o + e \quad (4)$$

where $o$ denotes the unknown bad data vector with its entry $o(i)$ being non-zero only if $z(i)$ is a bad datum.

A robust SE method, with capability to compress bad measurements, may be formulated as[11],[21]-[23]:

$$\min \quad J(x,o) = \frac{1}{2}[z-h(x)-o]^T[z-h(x)-o] + \lambda\|o\|_1$$
$$\text{s.t.} \quad h_e(x) = z_e \quad (5)$$

where $\lambda$ is a positive parameter.

## III. ROBUST BILINEAR STATE ESTIMATION

Note that (5) is a non-convex problem and very difficult to solve. However, the non-convexity of the measurement equations can be handled by using bilinear state estimation [18].

### A. First Linear Stage

For every branch connecting buses $i$ and $j$, we may define the following pair of variables:

$$K_{ij} = V_i V_j \cos\theta_{ij} \quad (6)$$

and

$$L_{ij} = V_i V_j \sin\theta_{ij} \quad (7)$$

where $\theta_{ij} = \theta_i - \theta_j$.

In addition, the squared voltage magnitude vector

$$U_i = V_i^2 \quad (8)$$

is included in the intermediate state vector $y$, which consists of $2b+N$ variables, i.e.,

$$y = \{U_i, K_{ij}, L_{ij}\} \quad (9)$$

Conventional measurement equations can then be linearly expressed in terms of $y$, i.e.,

$$P_i^m = \sum_{j\in i}P_{ij} + \varepsilon_P = \sum_{j\in i}^{j\neq i}[(g_{sh,i}+g_{ij})U_i - g_{ij}K_{ij} - b_{ij}L_{ij}] + \varepsilon_P \quad (10)$$

and

$$Q_i^m = \sum_{j\in i}Q_{ij} + \varepsilon_Q = \sum_{j\in i}^{j\neq i}[-(b_{sh,i}+b_{ij})U_i - g_{ij}L_{ij} + b_{ij}K_{ij}] + \varepsilon_Q \quad (11)$$

$$(V_i^2)^m = U_i + \varepsilon_U \quad (12)$$

The optimization problem in this stage amounts to the following compact form:

$$(y^*, o^{f*}) = \arg\min J^f(y, o^f)$$
$$= \arg\min \frac{1}{2}[z-By-o^f]^T[z-By-o^f] + \lambda\|o^f\|_1 \quad (13)$$
$$\text{s.t.} \quad Ey = z_e$$

where $o^f$ represents the bad data vector in the first linear stage, and $E$ is the counterpart of $B$ for exact-injection measurements.

### B. Intermediate Nonlinear Transformation

The intermediate vector $u$ is composed similarly to $y$, and contains $2b+N$ variables; it is defined as follows:

$$\alpha_i = 2\ln V_i, \quad (14)$$
$$\theta_{ij} = \theta_i - \theta_j \quad (15)$$

and

$$\alpha_{ij} = \alpha_i + \alpha_j . \quad (16)$$

The elements of vector $u$ are given by

$$u = \{\alpha_i, \theta_{ij}, \alpha_{ij}\} \quad (17)$$

and can be explicitly expressed in terms of $y$ as follows:

$$\alpha_i = \ln U_i, \quad (18)$$
$$\alpha_{ij} = \ln(K_{ij}^2 + L_{ij}^2) \quad (19)$$

and

$$\theta_{ij} = \arctan(\frac{L_{ij}}{K_{ij}}) . \quad (20)$$

These three equations (18)-(20) constitute the nonlinear transformation

$$u^* = f_u(y^*) . \quad (21)$$

The transformation of weighting matrices in the two linear stages mentioned in [18] can be replaced by some approximation methods in distributed manner. But numerical results have shown that when using robust SE model, (13) and (27) is satisfactory for engineering practice without the transformation of weighting matrices.

### C. Second Linear Stage

The terms $u$ and $x$ can be expressed in blocked form as follows:

$$u^* = [\alpha^T, \alpha_b^T, \theta_b^T]^T \quad (22)$$

and

$$x = [\alpha^T, \theta^T]^T \quad (23)$$

where the sub-index $b$ represents the set of branch variables.

When bus voltage measurements from PMUs are available, the phase angle can be directly incorporated into $u^*$ as well:

$$u^* = [\alpha^T, \alpha_b^T, \theta^T, \theta_b^T]^T \quad (24)$$

The branch components of $u$ can be expressed in terms of $x$ as follows:

$$\alpha_b = |A^T|\alpha \quad (25)$$
$$\theta_b = A_r^T\theta \quad (26)$$

where $A$ is the well-known branch-to-node incidence matrix, and $A_r$ represents the reduced matrix obtained by eliminating the reference angle in $A$.

The following compact optimization problem must be solved at this stage:

$$\min \quad J^s(x, o^s) = \frac{1}{2}[u^* - Cx - o^s]^T[u^* - Cx - o^s] + \lambda\|o^s\|_1 \quad (27)$$

where $o^s$ is a vector containing the bad data in the second linear stage.

## IV. Efficient Fully Distributed Algorithm

In this section, we describe the efficient fully distributed algorithm used to solve the RBSE. The structure of the RBSE problem is exploited to expedite the distributed algorithm by decomposing the two linear stages into independent calculations for each area, and deriving closed-form solution for sub-problem in each iteration of the ADMM.

### A. Decomposition in the First Linear Stage

Consider an inter-connected system consisting of $R$ areas. The $a$-th area supervises bus set $N_a$, internal branch set $E_a$. Measurements $z_a$ and $o_a^f$ are sub-vectors of $z$ and $o^f$, respectively, according to the partition. Due to the coupling of tie-lines between areas in the optimization problem in (13), the branch variables $K_{ij}$ and $L_{ij}$ over the tie-lines have to be shared by neighboring areas connected by them. Let $y_a = \{U_{a,i} \mid \forall i \in N_a\} \cup \{K_{a,ij}, L_{a,ij} \mid \forall (i,j) \in E_a \cup \Gamma_{a,b}\}$ denote the local copies of $y$ in area $a$ respectively. Using the variable splitting technique, the first linear stage can be transformed equivalently into following form:

$$\min \quad J^f(y, o^f) = \sum_{a=1}^{R} J_a^f(y_a, o_a^f) = \tag{28}$$
$$\sum_{a=1}^{R} [\frac{1}{2}(z_a - B_a y_a - o_a^f)^T (z_a - B_a y_a - o_a^f) + \lambda \|o_a^f\|_1]$$

s.t.
$$E_a y_a = z_{e,a}, \quad \forall a \tag{29}$$

$$K_{a,ij} = K_{ij}, \quad \forall (i,j) \in \Gamma_{a,b}, \forall b \in \Delta_a, \forall a. \tag{30}$$
$$L_{a,ij} = L_{ij}$$

Particular attention should be paid to consensus constraints(30), which implies coupling across areas over tie-lines. Fig. 1 demonstrates an example of coupling across areas. Different areas are coupled in a way of consensus to the global state variables, as shown in Fig. 1.

The ADMM described in[11],[23]-[25] is employed to decompose problem described by (28)-(30) per area by relaxing all the coupling constraints. The corresponding augmented Lagrangian function is defined as follows:

$$L^f(\{y_a\},\{o_a^f\},y,\eta,\gamma) = \sum_{a=1}^{R} L_a^f(y_a, o_a^f, y, \eta_a, \gamma_a)$$
$$= \sum_{a=1}^{R} \Bigg\{ J_a^f(y_a, o_a^f) + \sum_{b \in \Delta_a} \sum_{(i,j) \in \Gamma_{a,b}} [\eta_{a,ij}(K_{a,ij} - K_{ij}) \tag{31}$$
$$+ \gamma_{a,ij}(L_{a,ij} - L_{ij}) + \frac{\rho^f}{2}(\|K_{a,ij} - K_{ij}\|_2^2 + \|L_{a,ij} - L_{ij}\|_2^2)] \Bigg\}$$

where $\eta_{a,ij}$ and $\gamma_{a,ij}$ are the Lagrangian multipliers corresponding to constraint (30), $\rho^f \in R^+$ is a constant penalty parameter. Note that the global Lagrangian was decoupled spatially in (31).

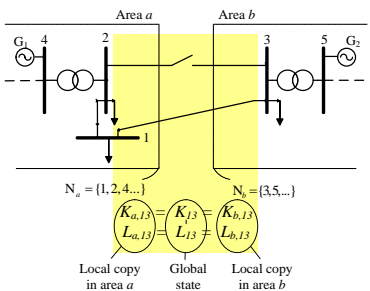

Fig. 1 Illustrative example of area decoupling in the first linear stage. Yellow zone indicates coupling over tie-lines between areas.

By applying the ADMM technique, the problem in (28)-(30) can be split into $R$ independent problems, i.e., one per area. The $t$-th iteration of the distributed algorithm can be written as follows:

$$y_a^{t+1} = \arg\min_{y_a} L_a^f\left(y_a, o_a^{f,t}, y^t, \eta_a^t, \gamma_a^t\right), \quad \forall a, \tag{32}$$
$$s.t. \quad E_a y_a = z_{e,a}$$

$$\{o_a^{f,t+1}\}, y^{t+1}\} = \arg\min_{\{o_a^f\}, y} L\left(\{y_a^{t+1}\}, \{o_a^f\}, y^t, \eta_a^t, \gamma_a^t\right) \tag{33}$$

and

$$\eta_{a,ij}^{t+1} = \eta_{a,ij}^t + \rho^f (K_{a,ij}^{t+1} - K_{ij}^{t+1})$$
$$\gamma_{a,ij}^{t+1} = \gamma_{a,ij}^t + \rho^f (L_{a,ij}^{t+1} - L_{ij}^{t+1}), \quad \forall (i,j) \in \Gamma_{a,b}, \forall b \in \Delta_a, \forall a. \tag{34}$$

However, it is clear that updating global state vector $y$ in (33) requires central coordination. Therefore, a distributed algorithm that does not require central coordination is devised here by eliminating the global vector $y$. The algorithm is further accelerated by deducing the closed-form solution in each ADMM iteration. And the cycles in (32)-(34) are equivalent to the following iterations (a sketch of proof is available in appendix):

$$y_a^{t+1} = \hat{G}_{B,a}^{-1}(I - E_a^T \hat{B}_a)[B_a^T (z_a - o_a^{f,t}) + \rho^f \hat{y}_a^t] + \hat{B}_a^T z_{e,a}, \tag{35}$$

$$o_a^{f,t+1} = [z_a - B_a y_a^{t+1}]_\lambda^+, \tag{36}$$

$$\bar{K}_{a,ij}^{t+1} = \frac{1}{2}(K_{a,ij}^{t+1} + K_{b,ij}^{t+1})$$
$$\bar{L}_{a,ij}^{t+1} = \frac{1}{2}(L_{a,ij}^{t+1} + L_{b,ij}^{t+1}), \quad \forall (i,j) \in \Gamma_{a,b}, \forall b \in \Delta_a \tag{37}$$

and

$$\hat{K}_{a,ij}^{t+1} = \hat{K}_{a,ij}^t + 2\bar{K}_{a,ij}^{t+1} - \bar{K}_{a,ij}^t - K_{a,ij}^{t+1}$$
$$\hat{L}_{a,ij}^{t+1} = \hat{L}_{a,ij}^t + 2\bar{L}_{a,ij}^{t+1} - \bar{L}_{a,ij}^t - L_{a,ij}^{t+1}, \quad \forall (i,j) \in \Gamma_{a,b}, \forall b \in \Delta_a \tag{38}$$

where $\hat{G}_{B,a}$ is the constant augmented gain matrix

$$\hat{G}_{B,a} = B_a^T B_a + \rho^f I, \tag{39}$$

$\hat{B}_a$ is an constant matrix

$$\hat{B}_a = (E_a \hat{G}_{B,a}^{-1} E_a^T)^{-1} E_a \hat{G}_{B,a}^{-1}, \tag{40}$$

and $[\cdot]_\lambda^+$ denotes the thresholding operator

$$[\xi_a(l)]_\lambda^+ = \begin{cases} \xi_a(l) + \lambda, & \xi_a(l) < -\lambda \\ \xi_a(l) - \lambda, & \xi_a(l) > \lambda \\ 0, & \text{otherwise} \end{cases} \tag{41}$$

which is required for the $l$-th entry in (36). $\hat{y}_a^t = \{\hat{U}_{a,i}^t, \hat{K}_{a,ij}^t, \hat{L}_{a,ij}^t\}$ is a sparse auxiliary vector with the same structure of $y_a^t = \{U_{a,i}^t, K_{a,ij}^t, L_{a,ij}^t\}$, except that only branch variables over tie-lines are defined as (38) while other elements remain zero in $\hat{y}_a^t$.

In the $t$-th iteration, the primal residual vector can be defined as

$$r^{f,t} = \frac{1}{2}\{|K_{a,ij}^t - K_{b,ij}^t|, |L_{a,ij}^t - L_{b,ij}^t| \mid \forall (i,j) \in \Gamma_{a,b}, \forall b \in \Delta_a, \forall a\} \tag{42}$$

which quantifies the mismatch between the area and its neighbors at the border between them.

The dual residual vector is defined as

$$d^{f,t} = \{|\bar{K}_{a,ij}^t - \bar{K}_{a,ij}^{t-1}|, |\bar{L}_{a,ij}^t - \bar{L}_{a,ij}^{t-1}| \mid \forall (i,j) \in \Gamma_{a,b}, \forall b \in \Delta_a, \forall a\} \tag{43}$$

which describes the stability of the iteration process.

The convergence of the first stage can be checked by a sufficiently small residual [23]

$$\delta^{f,t} = \left\| \begin{matrix} \boldsymbol{r}^{f,t} \\ \boldsymbol{d}^{f,t} \end{matrix} \right\|_\infty . \quad (44)$$

### B. Local Transformation in the Intermediate Nonlinear Transformation

Similar to the partition of $\boldsymbol{y}$, the intermediate vector $\boldsymbol{u}$ can be also separated into sub-vectors for different areas, i.e.,

$$\boldsymbol{u}_a = \{\alpha_{a,i} \mid \forall i \in \mathrm{N}_a\} \cup \{\theta_{a,ij}, \alpha_{a,ij} \mid \forall (i,j) \in \mathrm{E}_a \cup \Gamma_{a,b}\}. \quad (45)$$

As shown in previous section, the local vectors $\boldsymbol{u}_a, \boldsymbol{y}_a$ between different areas overlap over tie-lines for the purpose of convergence of the first stage. However, since the mismatch over tie-lines is sufficiently small after the convergence (as shown in (42)), the overlapping variables over tie-line are not necessary any more.

To avoid redundancy, branch variables $K_{a,ij}, L_{a,ij}, \alpha_{a,ij}, \theta_{a,ij}$ over tie-lines are uniquely assigned to one area (e.g., the area with a smaller index) rather than shared by two adjacent areas. To this end, the set of tie-lines is revised as follows:

$$\hat{\Gamma}_{a,b} = \begin{cases} \Gamma_{a,b}, & \text{if } a < b, \\ \varnothing, & \text{otherwise} \end{cases}, \quad (46)$$

and the local vectors have also been modified accordingly:

$$\begin{aligned} \tilde{\boldsymbol{y}}_a &= \{U_{a,i} \mid \forall i \in \mathrm{N}_a\} \cup \{K_{a,ij}, L_{a,ij} \mid \forall (i,j) \in \mathrm{E}_a \cup \hat{\Gamma}_{a,b}\} \\ \tilde{\boldsymbol{u}}_a &= \{\alpha_{a,i} \mid \forall i \in \mathrm{N}_a\} \cup \{\theta_{a,ij}, \alpha_{a,ij} \mid \forall (i,j) \in \mathrm{E}_a \cup \hat{\Gamma}_{a,b}\} \end{aligned} \quad (47)$$

Now that local vectors are completely decoupled, the nonlinear transformation can also be implemented in a fully distributed fashion:

$$\alpha_{a,i} = \ln U_{a,i} , \quad (48)$$

$$\alpha_{a,ij} = \ln(K_{a,ij}^2 + L_{a,ij}^2) \quad (49)$$

and

$$\theta_{a,ij} = \arctan\left(\frac{L_{a,ij}}{K_{a,ij}}\right) . \quad (50)$$

Note that neither the input data nor the output data are coupled across buses, as shown in Fig. 2. And therefore, the local transformation can be performed at each area in parallel.

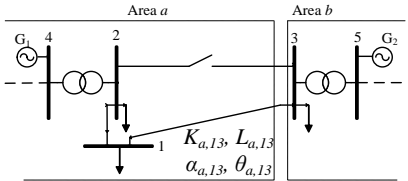

Fig. 2 Illustrative diagram of area decoupling in nonlinear transformation (suppose $a<b$).

### C. Decomposition in the Second Linear Stage

In this stage, the input "measurements" $\boldsymbol{u}$ have been separated into non-overlapping local "measurements" $\{u_a\}$, but the branch measurements, e.g., $\alpha_{a,13}, \theta_{a,13}$, are related to the other end of the tie-line, e.g. bus 3, that lie outside area $a$. To tackle this challenge, the boundary buses in areas with larger index, e.g., area $b$, have to be shared by its neighboring areas, e.g., area $a$, and the corresponding bus set is defined:

$$\hat{\mathrm{N}}_a^{BB} = \{i \mid (i,j) \in \Gamma_{a,b}, a > b\} \cup \{j \mid (i,j) \in \Gamma_{a,b}, a < b\} \quad (51)$$

Since the state variables in this stage are all nodal variables, the global state variables at each bus can be represented in vector form $\boldsymbol{x}_i = (\alpha_i, \theta_i)^T$, and local state variables $\boldsymbol{x}_a = \{\boldsymbol{x}_{a,i} \mid \forall i \in \mathrm{N}_a \cup \hat{\mathrm{N}}_a^{BB}\}$. The optimization problem in (27) can then be decomposed as follows:

$$\min \quad J^s(\boldsymbol{x}, \boldsymbol{o}^s) = \sum_{a=1}^R J_a^s(\boldsymbol{x}_a, \boldsymbol{o}_a^s) =$$

$$\sum_{a=1}^R [\frac{1}{2}(\tilde{\boldsymbol{u}}_a - \boldsymbol{C}_a \boldsymbol{x}_a - \boldsymbol{o}_a^s)^T (\tilde{\boldsymbol{u}}_a - \boldsymbol{C}_a \boldsymbol{x}_a - \boldsymbol{o}_a^s) + \lambda \|\boldsymbol{o}_a^s\|_1] \quad (52)$$

$$s.t. \quad \boldsymbol{x}_{a,i} = \boldsymbol{x}_i, \quad \forall i \in \hat{\mathrm{N}}_a^{BB}, \forall a . \quad (53)$$

Different from those in the first stage, the consensus constraints (53) reflect coupling at boundary buses. It can be observed by comparing the illustrative diagram Fig. 3 and Fig. 1 that inter-regional coupling in this stage has transferred from branch variables over tie-lines to nodal variables at boundary buses.

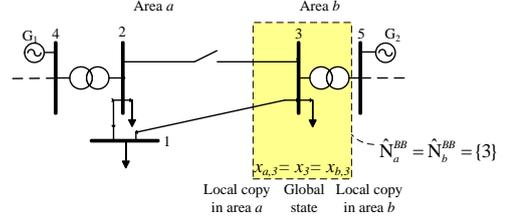

Fig. 3 Illustrative example of area decoupling in the second linear stage. Yellow zone indicates coupling over boundary buses.

The problem (52)-(53), which is similar to the problem in the first stage apart from the zero injection constraints(29), can also be solved by the distributed ADMM solver. The procedure is tantamount to the following iterations:

$$\boldsymbol{x}_a^{t+1} = \hat{\boldsymbol{G}}_{C,a}^{-1}(\boldsymbol{C}_a^T(\tilde{\boldsymbol{u}}_a - \boldsymbol{o}_a^{s,t}) + \rho^s \hat{\boldsymbol{x}}_a^t) , \quad (54)$$

$$\boldsymbol{o}_a^{s,t+1} = [\tilde{\boldsymbol{u}}_a - \boldsymbol{C}_a \boldsymbol{x}_a^{t+1}]_\lambda^+ , \quad (55)$$

$$\overline{\boldsymbol{x}}_{a,i}^{t+1} = \frac{1}{m_i} \sum_{a \in \mathrm{M}_i} \boldsymbol{x}_{a,i}^{t+1}, \quad \forall i \in \hat{\mathrm{N}}_a^{BB} \quad (56)$$

and

$$\hat{\boldsymbol{x}}_{a,i}^{t+1} = \hat{\boldsymbol{x}}_{a,i}^t + 2\overline{\boldsymbol{x}}_{a,i}^{t+1} - \overline{\boldsymbol{x}}_{a,i}^t - \boldsymbol{x}_{a,i}^{t+1}, \quad \forall i \in \hat{\mathrm{N}}_a^{BB} \quad (57)$$

where $\hat{\boldsymbol{G}}_{C,a}$ is the constant augmented gain matrix

$$\hat{\boldsymbol{G}}_{C,a} = \boldsymbol{C}_a^T \boldsymbol{C}_a + \rho^s \boldsymbol{I} ; \quad (58)$$

$\overline{\boldsymbol{x}}_a^t = \{\overline{\alpha}_{a,i}^t, \overline{\theta}_{a,i}^t\}$ and $\hat{\boldsymbol{x}}_a^t = \{\hat{\alpha}_{a,i}^t, \hat{\theta}_{a,i}^t\}$ are sparse auxiliary vectors with the same structure of $\boldsymbol{x}_a^t = \{\alpha_{a,i}^t, \theta_{a,i}^t\}$, except that only nodal variables that lie in $\hat{\mathrm{N}}_a^{BB}$ are defined as (56) and (57), while other elements remain zero in $\overline{\boldsymbol{x}}_a^t$ and $\hat{\boldsymbol{x}}_a^t$. $\mathrm{M}_i$ denotes the set of indices of areas that contain bus $i$ in its extended boundary bus set $\hat{\mathrm{N}}_a^{BB}$, i.e.,

$$\mathrm{M}_i = \{a \mid i \in \hat{\mathrm{N}}_a^{BB}\} \quad (59)$$

and $m_i$ its cardinality. Residual in the second stage $\delta^{s,t}$ can be defined in the same way as that in the first stage.

Note that in both linear stages, only communication among neighboring areas is required. The data required to exchange are the states of tie-lines or boundary buses, which amount to only a few float data for each area.

## V. SIMULATION RESULTS

The validity of centralized BSE has been discussed previously in[16][18]. In this section, three numerical experiments were conducted on interconnected test systems of different scales to examine the performance of D-RBSE. The first experiment was carried out on a two-area IEEE 14-bus system to illustrate the solution process, and to verify the solution quality in detail. The second test was performed using a three-area IEEE 118-bus system to demonstrate statistical accuracy of D-RBSE in quantities of scenarios Bus 1 is set as reference bus for all the three systems. For the latter system, a full measurement set is configured. Configurations of test systems are listed in Table I and detailed data are all referred to [26].

TABLE I CONFIGURATIONS OF THREE TEST SYSTEMS

| System | Areas | Units | Int. Lines | Tie-lines |
|---|---|---|---|---|
| 14-Bus | 2 | 5 | 6 | 3 |
| 118-Bus | 3 | 54 | 174 | 12 |

The SE algorithms were developed in Matlab R2013b using sparse matrix representations, and the simulations were carried out using a personal computer with an Intel Core i3-370M processor running at 2.4 GHz (4 GB RAM).

Measurement noise is simulated as independent zero-mean Gaussian with standard deviation 0.004 p.u. and 0.002 p.u. for power measurements and voltage magnitude measurements, respectively[16]. Bad data are simulated by adding Gaussian-distributed errors with a very large standard deviation (100 times larger than that of measurement noise) to the corresponding true value. $\lambda$ is empirically set as 1.34, $\rho^f = 1.0$, $\rho^s = 0.1$, and the tolerance for convergence was $\varepsilon = 5.0 \times 10^{-4}$.

To assess the accuracy of the estimated state, the performance metric here is the average absolute difference between the true value and estimated states:

$$S_V = \frac{1}{N} \|V - V^{true}\|_1 \quad (60)$$

and

$$S_\theta = \frac{1}{N-1} \|\theta - \theta^{true}\|_1 \quad (61)$$

where $V$ and $\theta$ denote the estimated results, while $V^{true}$ and $\theta^{true}$ represent the true value.

### A. Two-Area IEEE 14-Bus Interconnected System

A case study was carried out on an IEEE 14-bus interconnected system. As is shown in Fig. 4, the system is divided into two areas connected via three tie-lines. Measurements consist of voltage magnitudes at all buses, power flows across all branches (but "from" terminal only), power injections at all buses. Measurements are corrupted on branch power flow over tie-line (5, 6), power injection at boundary bus 5, and voltage magnitude at internal bus 14.

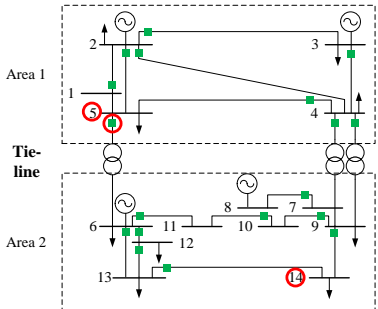

Fig. 4 Two-area IEEE 14-bus interconnected system. Branch (corrupted) measurements are depicted by green squares (red circles).

#### 1) Convergence.

To illustrate convergence of the ADMM iterations, residuals in both stages are depicted in Fig. 5(a). Clearly, the overlapping borders of two areas converged approximately with a linear rate in 20 iterations (18.3 msec), yielding a final estimation precision of ~$1.0 \times 10^{-4}$ in comparison to the true value.

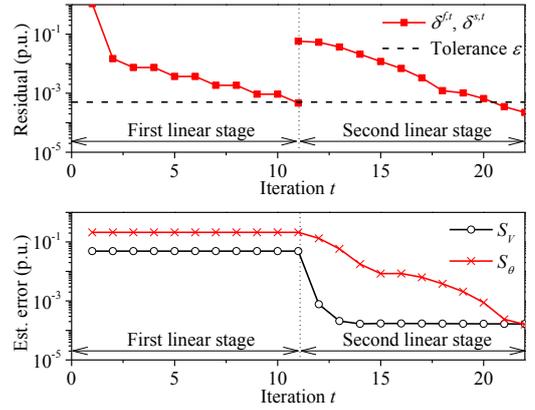

Fig. 5 Evolution of ADMM iterations in IEEE 14-bus system. (a) Residuals in both stages; and (b) estimation errors of magnitude and phase angle

#### 2) Accuracy analysis.

Fig. 6 further provides detailed comparison of the estimation results with different solution methods. True value of all states is depicted as purple "+". Influenced by the bad data, results of weighted least square (WLS) estimation, marked as black squares, stray far away from the true value. The convergence of D-RBSE guarantees that its results (blue "×") are identical to those of their counterparts (red circle) in the centralized RBSE. Thanks to the robust model, influence of bad data has been suppressed, and the results of both distributed and centralized RBSE are very close to true value. Table II describes the suppression of bad data in both internal and boundary regions.

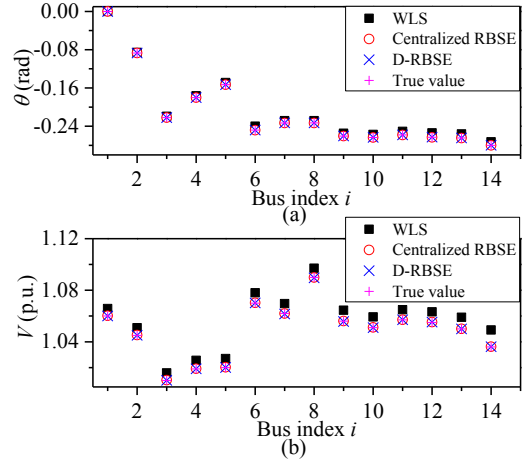

Fig. 6 Comparison of estimated states with different solution methods.

TABLE II SUPPRESSION OF BAD DATA IN THE IEEE 14-BUS SYSTEM

| Measurement | True value | Meas. value | WLS Est. results | WLS Est. error | D-RBSE Est. results | D-RBSE Est. error |
|---|---|---|---|---|---|---|
| Bus 5, $P_i$ | -0.0773 | 0.0387 | -0.0303 | 0.0470 | -0.0793 | 0.0020 |
| Bus 14, $V_i$ | 1.0360 | 1.1396 | 1.0492 | 0.0132 | 1.0362 | 0.0002 |
| Line (5,6), $P_{ij}$ | 0.4405 | 0.3137 | 0.4304 | 0.0101 | 0.4404 | 0.0001 |

### B. Three-Area IEEE 118-Bus Interconnected System

The IEEE 118-bus, with the same partition as Fig. 4 in[27], was tested next. Here, the bad data percentage is in the range of 0%-5%. For each scenario, the estimation errors $S_V$ and $S_\theta$ are averaged over 100 randomly-generated scenarios. In each scenario, corrupted measurements are randomly located in internal areas, boundary buses and tie-lines.

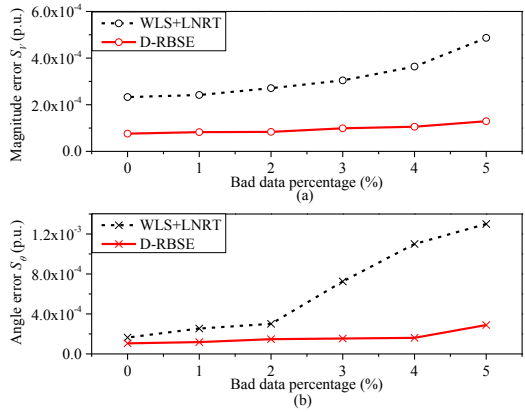

Fig. 7 Comparison of estimation errors between WLS+LNRT and D-RBSE in the IEEE 118-bus system versus the percentage of bad data for (a) $S_V$; and (b) $S_\theta$.

Fig. 7 shows state estimation errors of WLS with largest normalized residual tests (LNRT), depicted in red continuous lines, and D-RBSE (dashed black line). When there were no bad data, WLS, WLS+LNRT and D-RBSE exhibited almost identical results. In presence of 5% bad data, the performance of WLS deteriorates significantly, yielding an accuracy of ~$10^{-1}$, while both WLS+LNRT and D-RBSE can suppress the influence of bad data. However, D-RBSE performs slightly better than WLS+LNRT. Besides, the implementation of fully distributed LNRT is not straightforward.

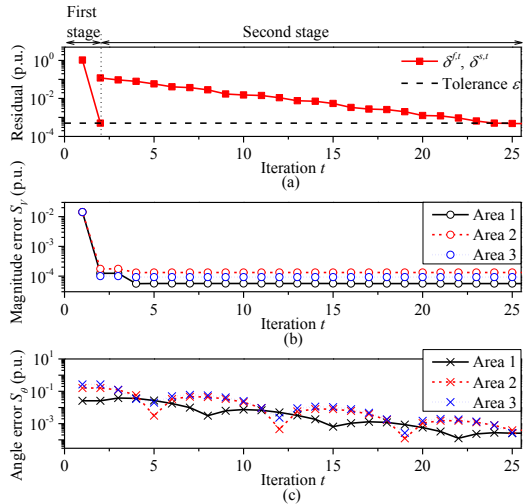

Fig. 8 Evolution of ADMM iterations in IEEE 118-bus system in presence of 5% bad data. (a) Residuals in both stages; (b) per area estimation error of magnitude and (c) phase angle.

When 5% of the measurements are corrupted, the corresponding convergence curve and error curves are plotted in Fig. 8. D-RBSE converges in 25 iterations with a final accuracy of ~$1.0 \times 10^{-4}$ in magnitude and ~$3.0 \times 10^{-4}$ in angle, while distributed SDP-based SE in [14] converged after about 20 iterations (215.6 msec) within an accuracy of ~$10^{-2}$ given bad data-free measurements (but please note that the metric is 2-norm there). Voltage magnitude converges within 5 iterations. But phase angle converges slower than magnitude, because in the second linear stage, there are almost no nodal information of phase angles except a single reference angle in area 1, which is transmitted to areas 2 and 3 in the form of boundary states. Due to fluctuating boundary angle, areas 2 and 3 converge in a rate slower than area 1. However, in case of installing a set of PMUs, the second stage improves significantly its efficiency with more nodal information of phase angle.

## VI. CONCLUSION

We described an extension of the centralized bilinear state estimation scheme to create a distributed robust bilinear state estimation method that is applicable to interconnected power systems with nonlinear measurements. In the two linear stages, the SE problem is decomposed into areas, where each area solves its own local SE problem with minimal data exchange among neighboring areas. The intermediate nonlinear transformation in between can be performed by every area independently without the need of inter-regional communication. Simulation results using benchmark networks with different scale show that D-RBSE is resilient and efficient even in the presence of bad data, with a very small communication overhead. The algorithm can be further accelerated by incorporating PMUs bus voltage measurements. This method can be extended for unbalanced distribution networks and it is the future works.